\pdfoutput=1

\documentclass[a4paper]{article}

\usepackage{INTERSPEECH2020}
\usepackage{multirow}
\usepackage{multicol}
\usepackage{hyperref}
\usepackage{footnote}
\usepackage{xcolor}
\usepackage{url}

\usepackage{subfigure}

\title{A Comparative Re-Assessment of Feature Extractors for Deep Speaker Embeddings}
\name{Xuechen Liu{$^1{}^,{}^2$}, Md Sahidullah{$^2$}, Tomi Kinnunen{$^1$}}
\address{
  {$^1$}School of Computing, University of Eastern Finland, Joensuu, Finland\\
  {$^2$}Universit\'{e} de Lorraine, CNRS, Inria, LORIA, F-54000, Nancy, France}
\email{xuechen.liu@inria.fr, md.sahidullah@inria.fr, tkinnu@cs.uef.fi}

\begin{document}

\maketitle
\begin{abstract}
Modern automatic speaker verification relies largely on deep neural networks (DNNs) trained on mel-frequency cepstral coefficient (MFCC) features. While there are alternative feature extraction methods based on phase, prosody and long-term temporal operations, they have not been extensively studied with DNN-based methods. We aim to fill this gap by providing extensive re-assessment of 14 feature extractors on VoxCeleb and SITW datasets. Our findings reveal that features equipped with techniques such as spectral centroids, group delay function, and integrated noise suppression provide promising alternatives to MFCCs for deep speaker embeddings extraction. Experimental results demonstrate up to 16.3\% (VoxCeleb) and 25.1\% (SITW) relative decrease in equal error rate (EER) to the baseline.

\end{abstract}

\section{Introduction} \label{sec:intro}
\emph{Automatic speaker verification} (ASV) \cite{Tomi_summaryasv2010} aims to determine whether two speech segments are from the same speaker or not. It finds applications in forensics, surveillance, access control, and home electronics.

While the field has long been dominated by approaches such as \emph{i-vectors} \cite{Dehak_ivector2011}, the focus has recently shifted to non-linear \emph{deep neural networks} (DNNs). They have been found to surpass previous solutions in many cases.

\begin{figure}[!t]
    \centering
    \includegraphics[width=60mm,scale=0.8]{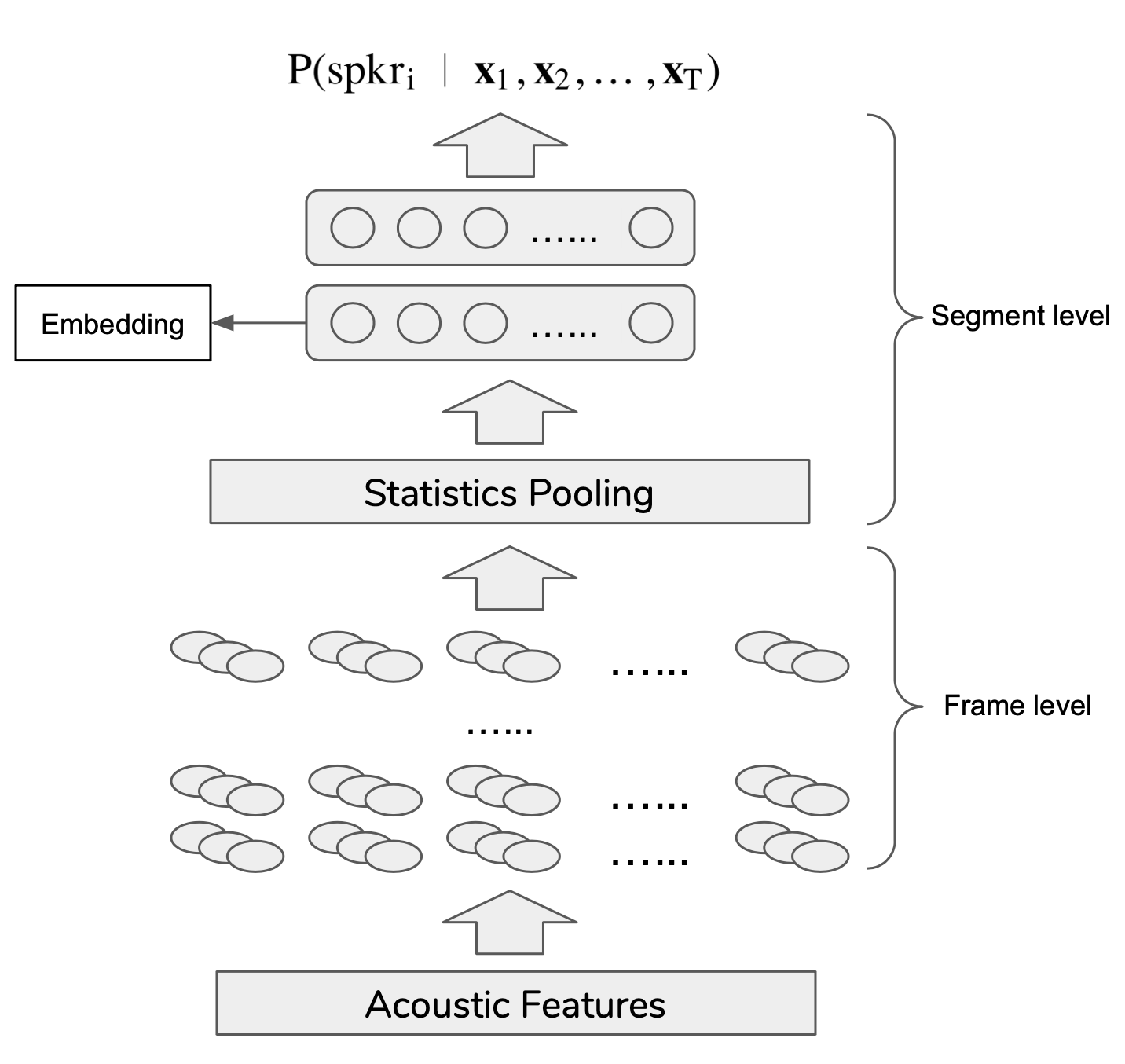}
    \vspace{-0.2cm}
    \caption{X-vector speaker embedding extractor \cite{Snyder_xvector2018}. Speaker embeddings are usually extracted from the first fully-connected layer after statistics pooling.}
    \label{fig:xvector}
\end{figure}

Representative DNN approaches include \emph{d-vector} \cite{Variani_dvector2014}, \emph{deep speaker} \cite{Baidu_deepspeaker2017} and \emph{x-vector} \cite{Snyder_xvector2018}. As illustrated in Figure \ref{fig:xvector}, DNNs are used to extract fixed-sized \emph{speaker embedding} from each utterance. These embeddings can then be used for speaker comparison with a back-end classifier. The network input and output consist of a sequence of acoustic feature vectors and a vector of speaker posteriors, respectively. The DNN learns input-output mapping through a number of intermediate layers, including temporal pooling (necessary for the extraction of fixed-sized embedding). 
A number of improvements to this core framework have been proposed, including \emph{hybrid frame-level layers} \cite{Snyder_extended_xvector2019}, use of \emph{multi-task learning} \cite{You_multitask_xvector2019} and alternative \emph{loss functions} \cite{Li_angular_softmax2018}, to name a few.
In addition, practitioners often use external data \cite{Ko_rir2017, Snyder_musan2015} to augment training data. This enforces the DNN to extract speaker-related attributes regardless of input perturbations.

While substantial amount of work has been devoted in improving DNN architectures, loss functions, and data augmentation recipes, the same cannot be said about acoustic features. There are, however, at least two important reasons to study feature extraction. First, data-driven models can only be as good as their input data --- the features. Second, in collaborative settings, it is customary to fuse several ASV systems. These systems should not only perform well in isolation, but be sufficiently \emph{diverse} as well. One way to achieve diversity is to train systems with different features. 

The acoustic features used to train deep speaker embedding extractors are typically standard \emph{mel-frequency cepstral coefficients} (MFCCs) or intermediate representations needed in MFCC extraction: raw spectrum \cite{Nagrani_vox1_2017}, mel-spectrum or mel-filterbank outputs. There are few exceptions where feature extractor is also learnt as part of the DNN architecture (\emph{e.g.} \cite{Ravanelli_sincnet2018}), 
although the empirical performance is often behind hand-crafted feature extraction schemes. This raises a question whether deep speaker embedding extractors might be improved by simple plug-and-play of other \emph{hand-crafted} feature extractors in place of MFCCs. Such methods are abundant in the past ASV literature \cite{ Tomi_multitaper2012, Rajan_apgdf2013, Kim_pncc2016}, and in the context of related tasks such as spoofing attack detection \cite{Sahid_feature_synthetic2015, Hanilci_feature_reply_attack2015}. An extensive study in the context of DNN-based ASV is however missing. 
Our study aims to fill this gap.

MFCCs are obtained from the power spectrum of a specific time-frequency representation, \emph{short-term Fourier transform} (STFT). MFCCs are therefore subjected to certain shortcomings of the STFT. They also lack specificity to the short-term phase of the signal. We therefore include a number of alternative features based on \textbf{short-term power spectrum} and \textbf{short-term phase}. Additionally, we also include \textbf{fundamental frequency} and methods that leverage from \textbf{long-term processing} beyond a short-time frame. Improvements over MFCCs are often motivated by robustness to additive noise, improved statistical properties, or closer alignment with human perception. The selected 14 features and their categorization, detailed below, is inspired from \cite{Sahid_feature_synthetic2015} and \cite{Hanilci_feature_reply_attack2015}. For generality, we carry experiments on two widely-adopted datasets, VoxCeleb \cite{Nagrani_vox1_2017} and speakers-in-the-wild (SITW) \cite{Mclaren_sitw2016}. To the best of our knowledge, this is the first extensive re-assessment of acoustic features for DNN-based ASV.

\section{Feature Extraction Methods} \label{sec:features}
In this section, we provide a comprehensive list of feature extractors with brief description for each method. Table \ref{tab:list_of_feats} summarizes the selected feature extractors along with their parameter settings and references to earlier ASV studies.

\begin{table*}[ht]
\centering
    \caption{List of feature extractors that are addressed in this study, with configuration details and references to exemplar earlier relevant studies on ASV. As mentioned in Section \ref{sec:intro} aside from MFCCs, previous works noted here are ones on conventional models.}
    \vspace{-0.2cm}
    {\footnotesize %
    \begin{tabular}{|c|c|c|c|c|}
    \hline
    Category & Feature (dim.) & Configuration details & Previous work on ASV \\
    \hline
    \multirow{5}{*}{\shortstack{Short-term magnitude \\ power spectral features}} & MFCC (30) & Baseline, No. of FFT coefficients=512 & \cite{Snyder_xvector2018, Snyder_extended_xvector2019} \\
        \cline{2-4}
        & CQCC (60) & CQCC\_v2.0 package\footnote{\url{http://www.audio.eurecom.fr/software/CQCC\_v2.0.zip}} & \cite{Todisco_artf_cqcc2016} \\ \cline{2-4}
        & LPCC (30) & LP order=30 & \cite{Jing_lpcc_asv2014} \\ \cline{2-4}
        & PLPCC (30) & LP order=30, bark-scale filterbank & \cite{Alam_multitaper_plpcc_2013} \\ \cline{2-4}
        & SCFC (30) & No. filters=30 & \multirow{2}{*}{\cite{Kua_scfc_scmc2010}} \\ \cline{2-3}
        & SCMC (30) & No. filters=30 & \\ \cline{2-4}
        & Multi-taper (30) & MFCC with SWCE windowing, no. tapers=8     & \cite{Tomi_multitaper2012, Alam_multitaper_plpcc_2013} \\ \cline{2-4}
    \hline
     \multirow{4}{*}{\shortstack{Short-term \\ phase spectral features}} & MGDF(30) & $\alpha = 0.4$, $\gamma = 0.9$, first 30 coeff. from DCT & \cite{Rajan_mgdf2009, Thiruvaran_phase_features2007} \\ \cline{2-4}
        & APGDF (30) & LP order=30 & \cite{Rajan_apgdf2013} \\ \cline{2-4}
        & CosPhase (30) & First 30 coeff. from DCT & - \\ \cline{2-4}
        & CMPOC (30) & $N=96$, First 30 coeff. from DCT & - \\ \cline{2-4} 
     \hline
     \multirow{2}{*}{\shortstack{Short-term features \\ with long-term processing}} 
     & MHEC (30) & No. of filters in Gammatone filter bank=20 & \cite{Sadjadi_mhec2015} \\ \cline{2-4}
     & PNCC (30) & First 30 coeff. from DCT & \cite{Wang_PNCC_asv2016}  \\ \cline{2-4}
    \hline
        Fundamental frequency features & MFCC+\textit{pitch} (33) & Kaldi pitch extractor, MFCC (30) with \textit{pitch} (3) & \cite{Adami_prosody_asv2007} \\
    \hline
    \end{tabular}}%
    \label{tab:list_of_feats}
 \end{table*}

\subsection{Short-term magnitude power spectral features}

\textbf{Mel frequency cepstral coefficients}. MFCCs are computed by integrating STFT power spectrum with overlapped band-pass filters on the mel-scale, followed by log compression and \emph{discrete cosine transform} (DCT). Following \cite{Tomi_summaryasv2010} a desired number of lower-order coefficients is retained. Standard MFCCs form our baseline features.

\textbf{Multi-taper mel frequency cepstral coefficients (Multi-taper)}. Viewing each short-term frame of speech as a realization of a random processes, the windowed STFT used in MFCC extraction is known to have high variance. To alleviate this, \emph{multi-taper} spectrum estimator is adopted \cite{Tomi_multitaper2012}. It uses several window functions (tapers) to obtain a low-variance power spectrum estimate, 
given by 
$\hat{S}(f)=\Sigma^{K}_{j=1}\lambda(j)|\Sigma^{N-1}_{t=0}w_{j}(t)x(t)e^{-i2{\pi}tf/N}|^{2}$.
Here, $w_{j}(t)$ is the $j$-th taper (window) 
and $\lambda(j)$ is its corresponding weight. The number of tapers, $K$, is an integer (typically between 4 and 8). There are a number of alternative taper sets to choose from: Thomson window \cite{Thomson_multitaper1982}, sinusoidal model (SWCE) \cite{Hansson_multitaper_swce2009} and multi-peak \cite{Hansson_multitaper_multipeak1995}. In this study, we chose SWCE. A detailed introduction of such spectrum estimator with experiments on conventional ASV can be found in \cite{Tomi_multitaper2012}.

\textbf{Linear prediction cepstral features}. An alternative to MFCC in terms of cepstral feature computation is from \emph{all-pole} \cite{Makhoul_lp_clasic1975} representation of signal. \emph{Linear prediction cepstral coefficients} (LPCCs) are derived from the linear prediction coefficients (LPCs) by a recursive operation \cite{Rabiner_basicspeech1993}. Similar method applies for perceptual LPCCs (PLPCCs) with applying a series of perceptual processing at primary stage \cite{Hermansky_plp_classic1990}.  

\textbf{Spectral subband centroid features}. Spectral subband centroid based features were introduced and investigated in statistical ASV \cite{Kua_scfc_scmc2010}. We consider two types of spectral centroid features: \emph{spectral centroid magnitude} (SCM) and \emph{subband centroid frequency} (SCF). They can be computed from weighted average of normalized energy of subband magnitude and frequency respectively. SCFs are then used directly as SCF coefficients (SCFCs) while log compression and DCT are performed for SCMs to obtain SCM coefficients (SCMCs). For more details one can refer to \cite{Kua_scfc_scmc2010}.

\textbf{Constant-Q cepstral coefficients (CQCCs)}. Constant-Q transform (CQT) was introduced in \cite{YoungBerg_cqt1978}. It has been applied in music signal processing \cite{Schorkhuber_cqt2010}, spoofing detection \cite{Todisco_cqcc2016} as well in ASV \cite{Delgado_icmc2016}. Different from STFT, CQT produces a time-frequency representation with variable resolution. The resulting CQT power spectrum is log-compressed and uniformly sampled, followed by DCT to yield CQCCs. Further details can be found in \cite{Todisco_cqcc2016}.

\subsection{Short-term phase features}

\textbf{Modified group delayed function (MGDF)}. MGDF was introduced in \cite{Murphy_mgdf2003} with application to phone recognition and further applied to speaker recognition \cite{Rajan_mgdf2009}. It is a parametric representation of the phase spectrum, defined as $\tau(k) = \mathrm{sign.} | X_{\text{R}}(k)Y_{\text{R}}(k)+Y_{\text{I}}(k)X_{\text{I}}(k)/(S(k))^{2\gamma} |^{\alpha}$, where $k$ is the frequency index; $X_{\text{R}}(k)$ and $X_{\text{I}}(k)$ are real and imaginary part of discrete Fourier transform (DFT) from speech samples $x(n)$; $Y_{\text{R}}(k)$ and $Y_{\text{I}}(k)$ are real and the imaginary parts of DFT of $nx(n)$. $\mathrm{sign.}$ is the the sign of $X_{\text{R}}(k)Y_{\text{R}}(k)+Y_{\text{I}}(k)X_{\text{I}}(k)$ while $\alpha$ and $\gamma$ are the control parameters; $S(k)$ is a smoothed magnitude spectrum. The cepstral-like coefficients which can be used as features are then obtained from function outputs by log-compression and DCT.

\textbf{All-pole group delayed function (APGDF)}. An alternative phase representation of signal was proposed for ASV in \cite{Rajan_apgdf2013}. The group delay function is computed by differentiating the unwrapped phase of all-pole spectrum. The main advantage of APGDF compared to MGDF is a fewer number of control parameters. 

\textbf{Cosine phase function (cosphase)}. Cosine of phase has been applied for spoofing attack detection \cite{Sahid_feature_synthetic2015, Wu_mgdf_synthetic2013}. The DFT-based unwrapped phase DFT is first normalized to $[-1, 1]$ using cosine operation, and then processed with DCT to derive the cosphase coefficients.

\textbf{Constant-Q magnitude–phase octave coefficients (CMPOCs)}. Unlike the previous DFT-based features, CMPOCs utilize CQT. The \emph{magnitude-phase spectrum} (MPS) from CQT is computed as $\sqrt{\text{ln}(|X(\omega))|^{2} + \phi(\omega)^{2}}$, where $X(\omega)$ and $\phi(\omega)$ denote magnitude and phase of CQT. Then, MPS is segmented according to the octave, and processed with log-compression and DCT to derive CMPOCs. The CMPOCs are studied so far for playback attack detection \cite{Yang_cmpoc2018}.

\subsection{Short-term features with long term processing}

We use the term `long-term processing' to refer methods that use information across a longer context of consecutive frames.

\textbf{Mean Hilbert envelope coefficients (MHECs)}. Proposed in \cite{Sadjadi_mhec2015} for i-vector based ASV, MHEC applies \emph{Gammatone} filterbanks on the speech signal. The output of each channel of the filterbank is then processed to compute temporal envelopes as $e_{s}(t,j)=s(t,j)+\hat{s}(t,j)$, where $s(t,j)$ is the so-called `analytical signal' and $\hat{s}(t,j)$ denotes its Hilbert transform \cite{Cohen_tfanalysis_textbook1995}. $t$ and $j$ represent time and channel index respectively. The envelopes are low-pass filtered, framed and averaged to compute energies. Finally, the energies are transformed to cepstral-like coefficients by log-compression and DCT. More details can be found in \cite{Sadjadi_mhec2015}.

\textbf{Power-normalized cepstral coefficients (PNCCs)}. To generate PNCCs input waveform is first processed by Gammatone filterbanks and fed into a cascade of non-linear time-varying operations, aimed at suppressing the impact of noise and reverberation. Mean power normalization is performed at the output of such operation series so as to minimize the potentially detrimental effect on amplitude scaling. Cepstral features are then obtained by power-law non-linearity and DCT. PNCCs have been applied to speech recognition \cite{Kim_pncc2016} as well as i-vector based ASV \cite{Wang_PNCC_asv2016}.

\subsection{Fundamental frequency features}

Aside from various type of features an initial investigation on the effect of harmonic information was conducted. For simplicity and comparability, the pitch extraction algorithm from \cite{Pegah_pitchasr2014} based on \emph{normalized cross correlation function} (NCCF) was employed to extract 3-dimensional pitch vectors. They are then appended to MFCCs.  In rest of the paper, we refer this feature as MFCC+\textit{pitch}.

\begin{table}[h]
    \centering
    \vspace{-0.2cm}
    \caption{Result of prior experiment on investigating dynamic features on Voxceleb1-E test set. Dimension of static part for all three cases were set to be 30. }
    {\footnotesize %
    \begin{tabular}{|c|c|c|}
    \hline
         Feature & EER(\%) & minDCF  \\ \hline
         MFCC & 4.65 & 0.5937  \\ \hline
         MFCC+$\Delta$ & 4.64 & 0.5517 \\ \hline
         MFCC+$\Delta\Delta$ & 4.77 & 0.5553 \\ \hline
    \end{tabular}}%
    \label{tab:prior_exp}
\end{table}{}

\begin{table}[h]
    \centering
    \vspace{-0.2cm}
    \caption{Result of different features and fusion systems on Voxceleb1-E test set and SITW development set (SITW-DEV).}
    {\footnotesize %
    \begin{tabular}{|c|c|c|c|c|}
    \hline
         & \multicolumn{2}{c|}{Voxceleb1-E} & \multicolumn{2}{c|}{SITW-DEV} \\ \hline
         Feature & EER(\%) & minDCF & EER(\%) & minDCF \\ \hline
         MFCC & 4.65 & 0.5937 & 8.12 & 0.8531 \\ \hline
         CQCC & 8.21 & 0.8310 & 9.43	& 0.9093 \\ \hline
         LPCC & 6.42 & 0.7129 & 9.39 & 0.9109 \\ \hline
         PLPCC & 7.06 & 0.7433 & 9.12 & 0.9178 \\ \hline
         SCFC & 6.56 & 0.7173 & 7.82	& 0.8530 \\ \hline
         SCMC & \textbf{4.57} & 0.5875 & 6.62 & 0.762 \\ \hline
         Multi-Taper & 4.84 & 0.5459 & 6.81 & 0.7776 \\ \hline
         MGDF & 7.73 & 0.7718 & 9.70 & 0.8878 \\ \hline
         APGDF & 5.96 & 0.6371 & 7.39 & 0.8449 \\ \hline
         cosphase & 6.03 & 0.6135 & 7.31 & 0.8436 \\ \hline
         CMPOC & 5.95 & 0.6758 & 7.62 & 0.8613 \\ \hline
         MHEC & 5.89 & 0.6777 & 7.66 & 0.8637 \\ \hline
         PNCC & 5.11 & 0.5659 & \textbf{6.08} & \textbf{0.7614} \\ \hline
         MFCC+\textit{pitch} & 4.67 & \textbf{0.5223} & 6.74 & 0.7983 \\ \hline\hline
         \fontsize{6}{4}\selectfont{MFCC+SCMC+Multi-taper} & \textbf{3.89} & 0.5396 & 6.58 & 0.7835 \\ \hline
         \scriptsize{MFCC+cosphase+PNCC} & 4.07 & 0.5103 & 6.24 & 0.7998 \\ \hline
    \end{tabular}
    }%
    \label{tab:main_exp}
\end{table}{}

\footnotetext{{\href{http://www.audio.eurecom.fr/software/CQCC\_v2.0.zip}{http://www.audio.eurecom.fr/software/CQCC\_v2.0.zip}}}

\section{Experiments} \label{sec:experiments}

\subsection{Datasets}
We conducted training of neural network on the \emph{dev} \cite{Nagrani_vox1_2017} part of Voxceleb1 consisting 1211 speakers. We used two evaluation sets, one for matched train-test condition and the other for relatively mismatched condition. First one was from the \emph{test} part of the same VoxCeleb1 dataset consisting 40 speakers, and the other one was from the \emph{development} part of SITW under ``core-core'' condition, consisting 119 speakers. The VoxCeleb1 evaluation consists of 18860 genuine trials and same number of imposter trials. On the other hand, the corresponding SITW partition has 2597 genuine and 335629 imposter trials. We will refer the two datasets as `\emph{Voxceleb1-E}' and `\emph{SITW-DEV}' respectively.

\subsection{Feature configuration}
Before being fed into feature extractors, we extracted all the features with a frame length of 25~ms and 10~ms shift. We apply Hamming \cite{Harris_windowing_classic1980} window in all cases except for the multi-taper feature. In Table \ref{tab:list_of_feats}, we describe the associated control parameters (if applicable) and the implementation details for each feature extractor. As for post-processing, we applied energy-based \emph{speech activity detection} (SAD) and utterance-level \emph{cepstral mean normalization} (CMN) \cite{Tomi_summaryasv2010} except for MFCC+\textit{pitch}, where the additional components contain \emph{probability of voicing} (POV).

\subsection{ASV system configuration}
To compare different feature extractors, we trained x-vector system for each of them, as illustrated in Figure \ref{fig:xvector}. We replicated the DNN configuration from \cite{Snyder_xvector2018}. We trained the model using data described above without any data augmentation. This will help to assess the inherent robustness of individual features. We extracted 512-dimensional speaker embedding for each test utterance. The embeddings were length-normalized and centered before being transformed using a 200-dimensional \emph{linear discriminant analysis} (LDA), followed by scoring with a \emph{probabilistic linear discriminant analysis} (PLDA) \cite{Ioffe_plda2006} classifier. 

\subsection{Evaluation}
The verification accuracy was measured by \emph{equal error rate} (EER) and \emph{minimum detection cost function} (minDCF) with target speaker prior $p=0.001$ and two costs $C_\text{FA}=C_\text{miss}=1.0$. \emph{Detection error trade-off} (DET) curves for all feature extraction methods are also presented. We used Kaldi\footnote{\href{https://github.com/kaldi-asr/kaldi}{https://github.com/kaldi-asr/kaldi}} for computing EER and minDCF. BOSARIS\footnote{\href{https://sites.google.com/site/bosaristoolkit/}{https://sites.google.com/site/bosaristoolkit/}} was called for DET illustration.

\begin{figure}[t]
\centering
\hspace*{-0.8cm}
\begin{subfigure}
\centering
  \includegraphics[width=0.45\paperwidth]{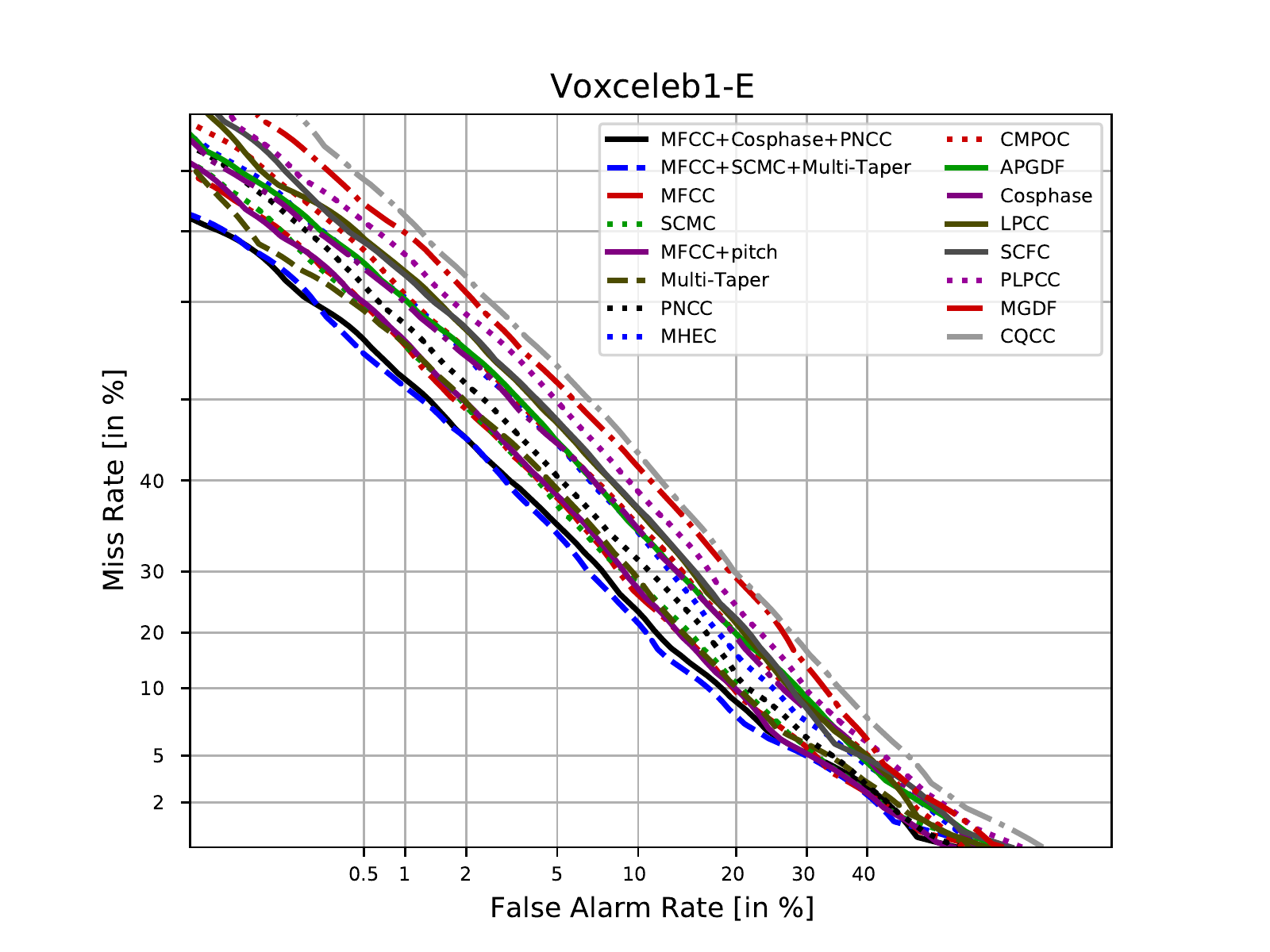}
  \label{fig:voxceleb1_det} 
\end{subfigure}
\hspace*{-0.8cm}
\begin{subfigure}
    \centering
  \includegraphics[width=0.45\paperwidth]{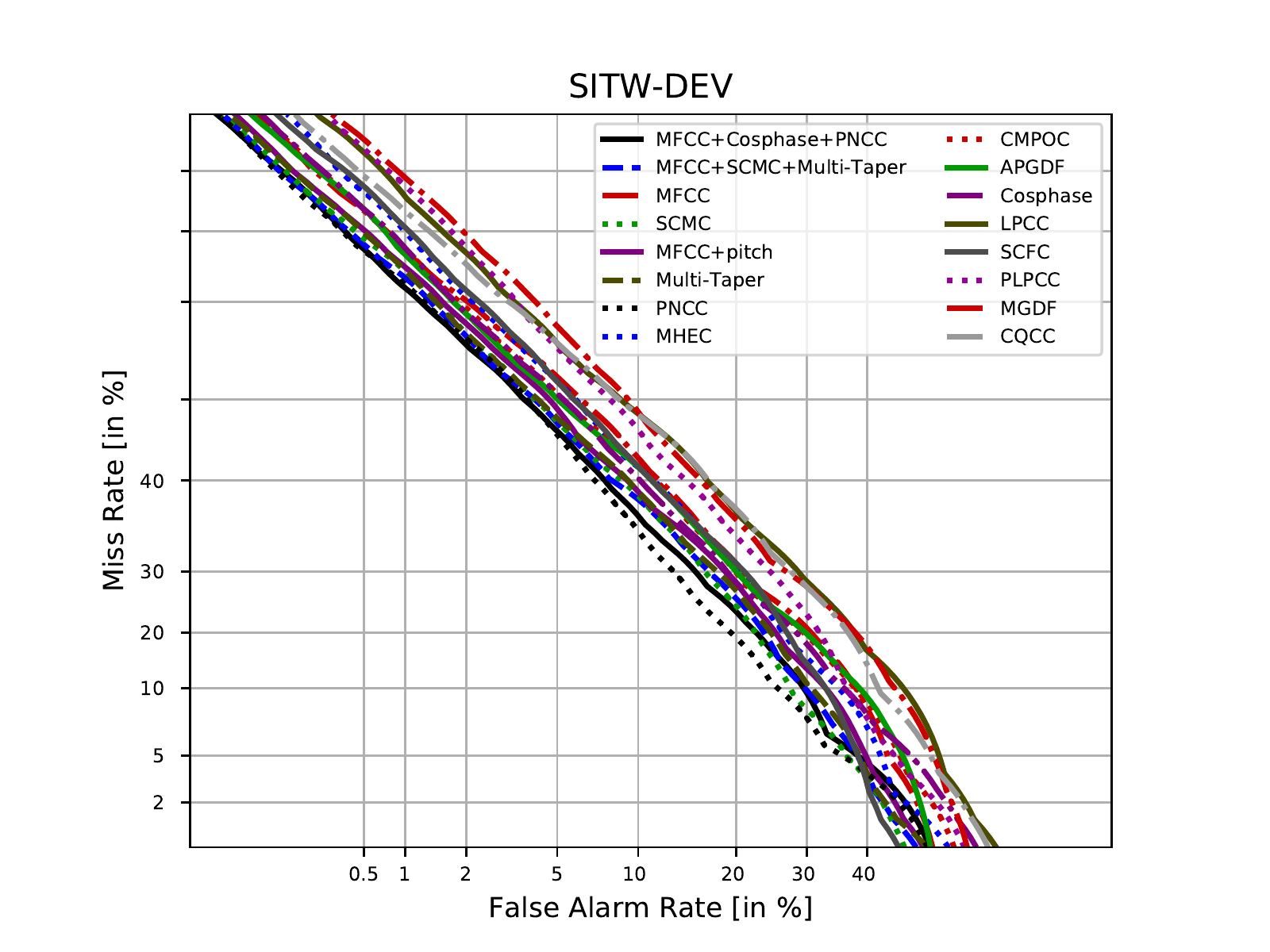}
  \label{fig:sitw_dev_set}
\end{subfigure}
\hspace*{\fill}
\caption{DET plots for evaluation sets. (top) Voxceleb1-E; (bottom) SITW-DEV. Best viewed in color.}
\label{fig:det_plots}
\end{figure}
\hspace*{\fill}

\section{Results} \label{sec:results}

We first conducted a preliminary experiment on investigating the effectiveness of dynamic features
with result reported in Table \ref{tab:prior_exp}, as a sanity check. We extended the baseline by adding delta and double-delta coefficients along with the static MFCCs. According to the table adding delta features did not improve performance.
This might be because the frame-level network layers already capture information across neighboring frames. In the remainder, we utilize static features only.

Table \ref{tab:main_exp} summarizes the results for both corpora.
In experiment of \emph{Voxceleb1-E}, we found that MFCCs outperform most of alternative features in terms of EER, with SCMCs as the only exception. This may indicate the effectiveness of information related to subband energies. However, SCFCs did not outperform SCMCs, which suggests that the subband magnitudes may be more important than their frequencies. Concerning phase spectral features, MGDFs were behind the other features. This might be due to sub-optimal control parameter settings. 
CMPOCs reached relatively 27.6\% lower EER than CQCCs, which highlights the effectiveness of phase features in CQT-based feature category. Moreover, while competitive EER and best minDCF can be observed from MFCC+\textit{pitch}, LPCCs and PLPCCs did not perform as good. This indicates the potential importance of explicit harmonic information. Such finding can be further found in \emph{SITW-DEV} results. Similar observation can be found from multi-taper MFCCs, which reclaims the efficacy of multi-taper windowing from conventional ASV.

Focusing more on \emph{SITW-DEV}, most competitive features include those from the phase and `long-term' categories. PNCCs reached best performance in both metrics, outperforming baseline MFCCs by 25.1\% relative in terms of EER. This might be due to the robustness-enhancing operations integrated in the pipeline, recalling that \emph{SITW-DEV} represents more challenging and mismatched data conditions. While not outperforming the baseline in \emph{Voxceleb1-E}, SCFCs yielded competitive numbers along with SCMCs, which further indicates usefulness of subband information. Best performance from cosphase under phase category reflects the advantage of cosine normalizer relative to group delay function. An additional benefit of cosphase over group delay features is that it has lesser number of control parameters.

Next, we addressed simple equal-weighted linear score fusion. We considered two sets of features: 1) MFCCs, SCMCs and Multi-taper; 2) MFCCs, cosphase and PNCCs. The former set of extractors share similar spectral operations while the latter cover more diverse speech attributes. Results are presented at the bottom of Table \ref{tab:main_exp}. In \emph{Voxceleb1-E}, we can see further improvement for both fused systems, especially for the first one which reached lowest overall EER, outperforming baseline by 16.3\% relatively. But under \emph{SITW-DEV} the best performance was still held by single system. This indicates that simple equal-weighted linear score-level fusion may be more effective for relatively matched conditions. 

Finally, the DET curves for all systems including fused ones are shown in Figure \ref{fig:det_plots}, which 
agrees with the findings in Table \ref{tab:main_exp}. 
Concerning \emph{Voxceleb1-E}, the two fusion systems are closer to the origin than any of the single systems in general, which corresponds to the indication above. Concerning SITW, PNCCs confirms its superior performance on \emph{SITW-DEV}, but from right-bottom both spectral centroid features are heading out, which may indicate their favor to systems that are less strict on false alarms.

\section{Conclusion}
This paper presents an extensive re-assessment of various acoustic feature extractors for DNN-based ASV systems. We evaluated them on Voxceleb1 and SITW, covering matched and unmatched conditions. We achieved improvements over MFCCs especially on SITW, which represents more mismatched testing condition. We also found alternative methods such as spectral centroids, group delay function, and integrated noise suppression can be useful for DNN system. For future work they thus shall be revisited and extended under more scenarios. Finally we gave an initial attempt on score-level fused systems with competitive performance, indicating the potential of such approach.

\section{Acknowledgements}
This work was partially supported by Academy of Finland (project 309629) and Inria Nancy Grand Est.

\bibliographystyle{IEEEtran}

\bibliography{mybib}

\end{document}